\documentclass[aps,aps,twocolumn]{revtex4}
\usepackage{graphicx}
\usepackage{dcolumn}
\usepackage{bm}
\usepackage{amsmath}
\usepackage{txfonts}
\usepackage[scaled]{helvet}
\usepackage{color}
\usepackage[breaklinks=true,colorlinks,citecolor=blue,linkcolor=blue,urlcolor=blue]{hyperref}
\usepackage{makecell}
\usepackage{booktabs}
\usepackage{threeparttable}

\begin{document}

\title{Proximity-induced diversified magnetic states and electrically-controllable spin polarization in bilayer graphene: Towards layered spintronics}
\author{Xuechao Zhai$^1$ and Yaroslav M. Blanter$^2$}
\affiliation{$^1$Department of Applied Physics and MIIT Key Laboratory of Semiconductor Microstructures and Quantum Sensing, Nanjing University of Science and
Technology, Nanjing 210094, China\\
$^2$Kavli Institute of Nanoscience, Delft University of Technology, 2628 CJ Delft, The Netherlands}

\date{\today}

\begin{abstract}
Compared to monolayer graphene, electrons in Bernal-stacked bilayer graphene (BLG) have an additional layer degree of freedom, offering a platform for developing {\it layered spintronics} with the help of proximity-induced magnetism. Based on an effective phenomenological model, we systematically study the effect of this magnetism on the spin-dependent band structure near the Fermi energy and identify the magnetic phases induced in BLG by proximity with magnets. We show that spin polarization can develop in BLG due to this proximity effect. This spin polarization depends strongly on the layer distribution of magnetism, and can always be controlled by gate voltage which shifts spin-dependent band edges and modifies the total band gap. We further show that the band spin polarization can be modified by the proximity-induced staggered sublattice potential. By taking full advantage of layer-dependent magnetism in BLG, we propose that spintronic devices such as a spin filter, a giant magnetoresistence device, and a spin diode can operate under fully electric control, which is easier than the common magnetic field control.
\end{abstract}

 \maketitle

\section{Introduction}

Bernal-stacked bilayer graphene (BLG) \cite{McKosh} is a van der Waals (vdW) layered two-dimensional (2D) system with the minimum possible layer number [Fig.~1(a)]. It provides an excellent platform to explore exotic phenomena, such as electrically-controllable band gap \cite{Neto}, topological valley Hall effect \cite{ShiYam}, or superconductivity in a twisted bilayer \cite{CaoFat}. The low-energy electrons in BLG are located near two Dirac points $K$ and $K'$ [Fig.~1(b)], referred to as valleys. Isolated ideal BLG is not magnetic \cite{McKosh}, however magnetism can be induced by external means, such as magnetic proximity effect \cite{GhiKav} or intercalation by magnetic atoms \cite{SokAve}, opening the route towards 2D spintronics applications.

Recently, externally induced magnetism in BLG caused considerable attention. In particular, it was theoretically shown that swapping of exchange coupling and spin-orbit coupling \cite{ZolGmi20} and layered antiferromagnetism \cite{ZhaiXu} can be realized using the vdW proximity effect in doubly-proximitized BLG systems. The proximity-induced magnetism results in a significant gate-tunable spin splitting near the Fermi energy, which can reach the order of 10~meV \cite{ZolGmi2018,ZolGmi20,ZhaiXu,ZolFab}. It has been experimentally demonstrated that the spin current is generated via electric and thermal means \cite{GhiKav} in BLG with proximity-induced magnetism. Despite this progress, systematic understanding of proximity-induced magnetic states in BLG is still missing.

In this Article, we make the first step towards such understanding by considering magnetism in BLG induced by magnetic materials applied to one or two sides of the BLG sample [Fig.~1(c)]. We only consider materials with out-of plane magnetization, which can induce ferromagnetic (FM), andiferromagnetic (AFM), or ferrimagnetic (FIM) order in BLG.  The out-of-plane magnetization we concentrate on favors ultrahigh-density data storage \cite{WelMos} and is usually superior to the canted or in-plane magnetization \cite{YouSan,Kharit}. Further considering the ultra-short range nature of the proximity effect \cite{ZolGmi2018,ZolGmi20,ZhaiXu,ZolFab}, we reasonably assume that the proximity-induced magnetism of two opposite layers of BLG is almost unaffected by the presence of another layer, as rigorously proven by previous first-principles calculations and fitting models \cite{ZolFab,ZhaiXu,ZolGmi2018,ZolGmi20}. This suggests that BLG is an excellent platform to engineer {\it layered spintronics}, which is proposed here as a separate branch of spintronics due to the independent variability and control of each layer of magnetism. The term {\em ``layered spintronics"} is different from that used in previous works \cite{BeraYus,TianGray,SunXiao}, which mention ``layered spintronics materials" but do not explore the possibility that magnetic structure of every layer can be modified. Encouragingly, the emerging of abundant ultrathin vdW materials \cite{LiuHuang,NovMish,ChoiYun,GoZha,LiYang} provides a rich soil for such modifications and thus for the development of {\it layered spintronics}.

Here, using an effective low-energy phenomenological model, we systematically study the diverse magnetic phases induced by the proximity effect in BLG and obtain the corresponding spin polarization near the Fermi energy. Our results indicate that the band spin polarization depends strongly on the layer distribution of magnetism and can always be controlled by gate voltage by turning on or off spin degeneracy and modulating spin-dependent band edges. We further demonstrate that the spin polarization can be modified by staggered sublattice potential originating from the proximity effect and illustrate how the electrically-tunable spin filtering, giant magnetoresistence, and spin diode effects can be visualized in the devices based on magnetic BLG junctions. The all-electrical control of our proposed devices is superior to the magnetic-field control of traditional devices \cite{Grunberg} because the operation is easier. Meanwhile, BLG can be easily integrated into heterostructures \cite{IsLew,GhiKav,SokAve}, and thus our results open new possibilities for developing low-dimensional spin-based devices by fully exploiting the layer degree of freedom.

The organization of this Article is as follows. In Sec.~II, we introduce the model Hamiltonian. In Sec.~III, we provide and discuss the results for layered magnetism. We conclude in Sec.~IV.

\begin{figure}
\centerline{\includegraphics[width=8.5cm]{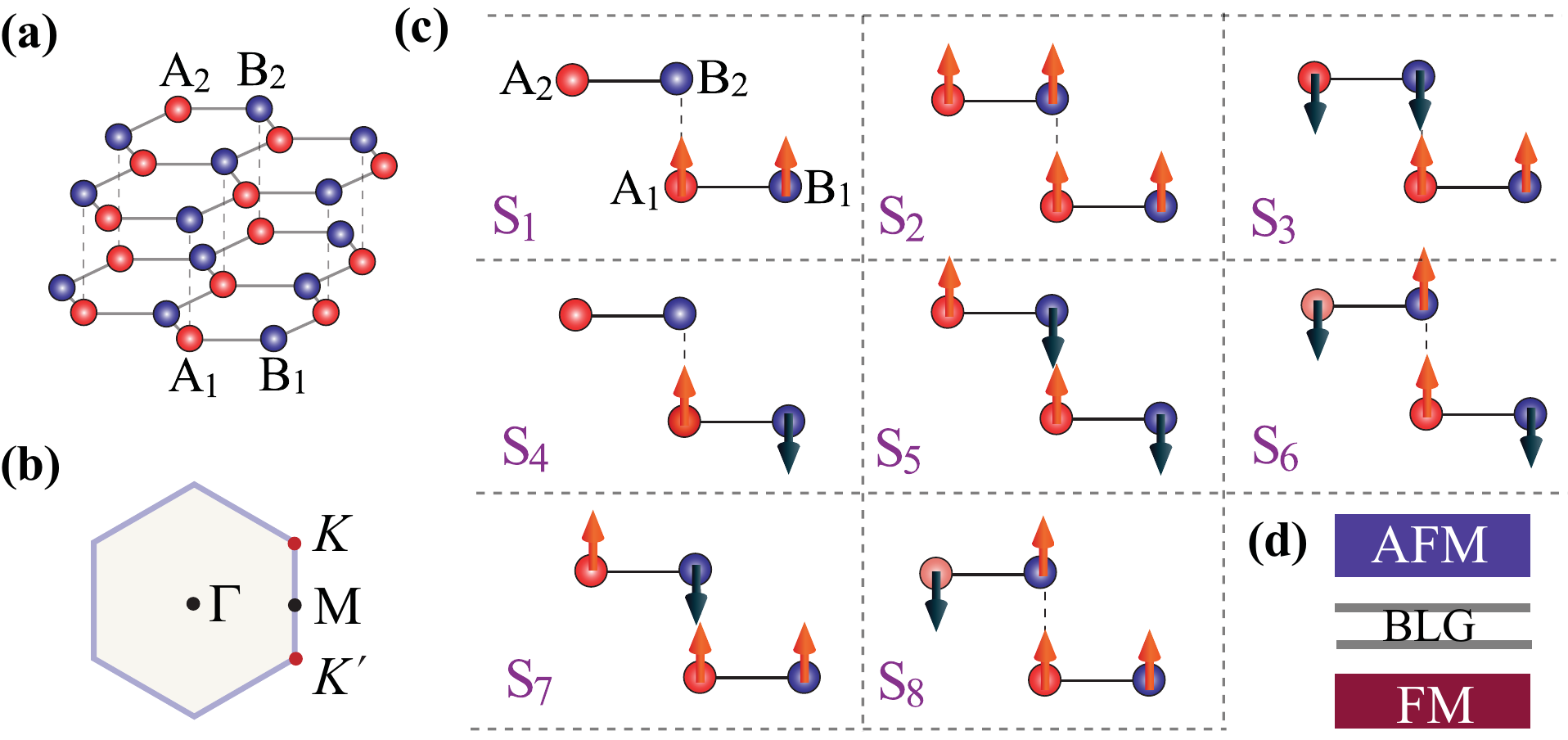}} \caption{(a) Bilayer graphene with AB stacking. A$_{i}$ and B$_{i}$ ($i=1,2$) represent two different sublattices in the $i$-th layer. (b) Brillouin zone. $\Gamma$, M, $K$ and $K'$ label four high-symmetry points. (c) Schematic of eight magnetic states labelled by ${\rm S}_j$ ($j=1\rightarrow8$) and distinguished by the orientation of magnetic moments of atoms. The atomic arrangement in ${\rm S}_2$ to  ${\rm S}_8$ is the same as that marked in S$_1$. Specifically, the atoms in the top layer have no magnetism for ${\rm S}_1$ and ${\rm S}_4$. (d) Realization of ${\rm S}_7$ or ${\rm S}_8$ by magnetic proximity effect from AFM and FM insulators on both sides of BLG.}
\end{figure}

\section{Model Hamiltonian}

Here, we consider BLG where one or both layers are in contact with (generally, different) layers of magnetic materials. In this Section, we construct the low-energy effective Hamiltonian of BLG, assuming that the exchange coupling is sufficiently strong, and the only role of the magnetic layers is to induce magnetism in one or both layers of BLG.

As shown in Fig.~1(a), BLG has four sublattices A$_1$, B$_1$, A$_2$ and B$_2$, where the subscript 1~(2) marks the bottom (top) layer. Besides charge, the electrons in BLG combine other multiple degrees of freedom \cite{McKosh,ZhaiBlanter2020}, including real spin, sublattice pseudospin (in each monolayer), layer pseudospin, and valley, which are described by the Pauli matrices ${\bm s}=(s_x, s_y, s_z)$, ${\bm \sigma}=(\sigma_x, \sigma_y, \sigma_z)$, ${\bm \tau}=(\tau_x, \tau_y, \tau_z)$, and the index $\xi$ [+1 for valley $K$ and -1 for valley $K'$, see Fig.~1(b)], respectively. The pseudospin $\bm \sigma$ is identical for two monolayers of BLG. Here, we employ ${\bm I_s}$, ${\bm I_\sigma}$ and ${\bm I_\tau}$, to denote the identity matrix in the spaces of $\bm s,\bm\sigma$ and $\bm\tau$, respectively. Taking $K$ ($K'$) as coordinate origin for $\xi=+1~(-1)$, the momentum is denoted by ${\bm p}=(p_x, p_y)$.

\begin{table} \label{T1}\footnotesize
\centering
\caption{Proximity-induced magnetism in graphene materials reported in the existing literature. Here, S$_1$ to S$_4$ correspond to the magnetic states of BLG in Fig.~1. As supplementary information, the hybrid systems composed of monolayer graphene (MLG) and magnetic insulators to realize the FM or AFM states are also listed. The abbreviation CGT stands for Cr$_2$Ge$_2$Te$_6$. Note that the hybrid systems BLG/Eu \cite{SokAve} and BLG/CrSBr \cite{GhiKav} are magnetized in plane and the others are magnetized out of the plane (without magnetic field). The magnetization is tunable by magnetic field \cite{GhiKav,SokAve}.}

\begin{threeparttable}
\begin{tabular}{lllll}
\toprule[1pt]
   State &\quad Hybrid systems \quad &\quad{Strength ($M$)} &\quad{Experiment}&\quad{Theory}\\

   \hline

S$_1$ &\quad BLG/CGT &\quad{8~meV} &\quad&\quad {\cite{ZolGmi2018}}\\
~~~~~~&\quad BLG/Eu&\quad{$\sim$1~meV}&\quad {\cite{SokAve}}\\
S$_{2(3)}$ &\quad CGT/BLG/CGT &\quad{3--10~meV} &\quad&\quad {\cite{ZhaiXu,ZolFab}}\\
~~~~~~&\quad CrI$_3$/BLG/CrI$_3$&\quad{8~meV}&\quad&\quad {\cite{CarSor}}\\
S$_4$ &\quad BLG/CrSBr &\quad{$\sim$20~meV} &\quad {\cite{GhiKav}}\\
\hline
FM&\quad MLG/EuO &\quad{$\sim$36~meV} &\quad {\cite{AveSok}}&\quad {\cite{YangHal}}\\
~~~~~~&\quad MLG/EuS &\quad{10--100~meV} &\quad {\cite{WeiLee}}&\quad {\cite{HalIbra}}\\
~~~~~~&\quad MLG/Y$_3$Fe$_5$O$_{12}$ &\quad{$\sim$80~meV} &\quad {\cite{WangTang}}&\quad {\cite{HalIbra}}\\
~~~~~~&\quad MLG/BiFeO$_3$ &\quad{30--70~meV} &\quad {\cite{WuSong}}&\quad {\cite{QiaoRen}}\\
~~~~~~&\quad MLG/CGT~(Strain) &\quad{2.8--13~meV} &\quad {\cite{KarpCum}}&\quad {\cite{ZhangZhao}}\\
~~~~~~&\quad MLG/CrI$_3$~(Strain) &\quad{65--150~meV} &\quad {\cite{NiuLu}}&\quad {\cite{ZhangZhao2}}\\
AFM&\quad MLG/MnPSe$_3$ &\quad{$\sim$0.1~meV} &\quad &\quad {\cite{HogFrank}}\\
~~~~~~&\quad MLG/CGT~(Twist)&\quad{$\sim$2~meV}&\quad &\quad {\cite{ZolFab2022}}\\
\bottomrule[1.0pt]
\end{tabular}
      \end{threeparttable}
\end{table}

Eight possible magnetic states ${\rm S}_1$ to ${\rm S}_8$ are shown in Fig.~1(c). In Table~I, we summarize the proximity-induced magnetism in graphene materials reported in the existing literature~\cite{GhiKav,SokAve,ZhaiXu,ZolGmi20,ZolGmi2018,ZolFab,
CarSor,AveSok,YangHal,WeiLee,HalIbra,WangTang,WuSong,QiaoRen,KarpCum,ZhangZhao,NiuLu,
ZhangZhao2,HogFrank,ZolFab2022}. It is seen that there have been reports of direct relevance for S$_1$ to S$_4$. For S$_5$ to S$_8$, no direct theoretical or experimental reports have been found yet, however we provide information on the single-interface proximity case, {\rm i.e.} monolayer graphene/magnetic insulator, as a reference. The configurations S$_5$ to S$_8$ become available if we select different magnets on the top and bottom sides. Taking S$_7$ and S$_8$ for example, they can be realized by contacting BLG with a ferromagnetic insulator from the bottom side and with an antiferromagnetic insulator from the top side. To study the electronic properties of the heterostructures, we use a simple assumption that the proximity effect only induces magnetism in BLG but does not change the band features of BLG, as proven previously \cite{SokAve,ZhaiXu,ZolGmi20,ZolGmi2018,ZolFab,AveSok,YangHal,WeiLee,HalIbra,HogFrank,ZolFab2022}. Specifically, the applied vertical pressure in heterostructures \cite{ZhangZhao,ZhangZhao2} is helpful to enhance the proximity-induced magnetism and avoid the introduction of extra orbital hybridization near the Fermi energy of graphene materials. Note that, since the magnets that induce magnetism from the top or bottom side of BLG basically only affect a single layer of BLG close to it, with another graphene layer being largely unaffected \cite{ZolFab,ZhaiXu,ZolGmi2018,ZolGmi20,ZolFab2022} under the ultra-short range proximity effect, the patterns ${\rm S}_1$ to ${\rm S}_8$ for the Bernal-stacking situation we consider, the most stable structure and mostly realized in experiments \cite{Neto,McKosh,ShiYam}, are also applicable to the relatively-rare AA-stacking situation \cite{RozSbo,LiuSue}. In the atomic basis $\{\psi_{A_1\uparrow}$, $\psi_{A_1\downarrow}$, $\psi_{B_1\uparrow}$,
$\psi_{B_1\downarrow}$, $\psi_{A_2\uparrow}$, $\psi_{A_2\downarrow}$, $\psi_{B_2\uparrow}$,
$\psi_{B_2\downarrow}\}$, we establish a generalized phenomenological Hamiltonian as follows,
\begin{equation}\label{H_EB}
\begin{split}
{\cal H}=&{\bm I_\tau}\otimes\upsilon(\sigma_xp_x+\xi\sigma_yp_y)\otimes{\bm I_s}
\\&+\frac{\gamma}{2}(\tau_x\otimes\sigma_x-\tau_y\otimes\sigma_y)\otimes{\bm I_s}\\
&+M\Sigma_{\tau,\sigma}\otimes s_z\\
&+\Delta\tau_{z,\eta'}\otimes\sigma_z\otimes{\bm I_s}\\
&+U\tau_z\otimes{\bm I_\sigma}\otimes{\bm I_s}\\
&+{\cal H}',\\
{\cal H}'=&\frac{\lambda_R}{2}\tau_{z,\nu}\otimes(\sigma_x\otimes s_y-\xi\sigma_y\otimes s_x)\\&+\lambda_I\xi \tau_{z,\chi}\otimes{\bm I_\sigma}\otimes s_z.
\end{split}
\end{equation}
The first term $H_\upsilon$ indicates the massless Dirac dispersion, where $\upsilon$ denotes the Fermi velocity in monolayer graphene. The second term $H_\gamma$ denotes the interlayer nearest-neighbor coupling, so that $H_\upsilon+H_\gamma$ describes the Hamiltonian for bare BLG \cite{Neto,McKosh,ShiYam}. The third term $H_M$ represents the proximity-induced magnetism. The fourth term $H_\Delta$ represents the staggered sublattice potential, which is zero for bare BLG but can appear when BLG is in proximity with other materials \cite{ZhangZhao,KarpCum,KriGol,SiPra,AvsOch,
BalKok,MarVar,FarTan,WangChe,AlsAsm,SwaOde}. The fifth term $H_U$ indicates the interlayer bias (tunable by gate voltage). Without loss of generality, the last term ${\cal H}'={\cal H}_R+{\cal H}_I$ presents the symmetry-breaking induced perturbations from Rashba-type ($\lambda_R$) and Ising-type ($\lambda_I$) spin-orbit couplings, which are the most common spin-orbit interactions induced in graphene-based interfaces \cite{ZolFab,ZhaiXu,ZolGmi2018,ZolGmi20,ZolFab2022}. To describe the dependence of magnetism and staggered sublattice potential on layer and sublattice, we parameterize
\begin{equation}
\Sigma_{\tau,\sigma}=
\left(\begin{array}{cccc}
1&0&0&0\\
0&\beta&0&0\\
0&0&\eta&0\\
0&0&0&\zeta\eta\\
\end{array}\right),~~
\tau_{z,\eta'}=
\left(\begin{array}{cc}
1&0\\
0&\eta'\\
\end{array}\right),
\end{equation}
where $\tau_{z,\eta}$ is defined in the $\bm \tau$ space, and the dimensionless parameters take the values
\begin{equation}
\{\beta,\zeta\}=\pm1,~\{\eta,\eta'\}\in[-1,1].
\end{equation}
Here, $\beta$, $\eta$, and $\zeta$ are three physical quantities reflecting the relative strength of local magnetism, for which the ratio of exchange splitting on sublattices $A_1$, $B_1$, $A_2$ and $B_2$ is $1:\beta:\eta:\zeta\eta$. By setting $\beta=\pm1$ and $\zeta=\pm1$, we assume for simplicity that the absolute values of magnetization in both sublatices of each monolayer of BLG are the same. The amplitude difference of magnetization between two opposite layers depends on the value of $\eta$. Because the proximity effect is short-range, $\beta$ and $\eta$ are two completely independent parameters. The index $\eta'$ depends on the coupling details between BLG and its proximity partners \cite{ZolGmi2018,ZolGmi20,ZhaiXu,ZolFab}. In Table~\ref{T2}, we show the magnetic type in the parameter space $\{\beta,\eta,\zeta\}$. Depending on the parameter values, the magnetism types of FM, AFM, FIM are distinguished. To avoid the complexity caused by too many parameters in the model~(\ref{H_EB}), we restrict ourselves below to the qualitative discussion of the effects of the spin-orbit coupling perturbation $\cal H'$.

\begin{table} \label{T2}\footnotesize
\centering
\caption{Magnetic type characterized by the parameters $\{\beta,\eta,\zeta\}$. Configurations S$_1$ to S$_8$ correspond to those in Fig.~1. The mark $``/"$ indicates that the value is not applicable.}

\begin{threeparttable}
\begin{tabular}{llc}
\toprule[1pt]
   State &\quad\qquad\qquad\qquad $\{\beta,\eta,\zeta\}$ &\quad\qquad\qquad\qquad{Magnetism Type}\\

   \hline

S$_1$ &\quad\qquad\qquad\qquad $\{1,0,/\}$ &\quad\qquad\qquad\qquad{FM}\\
S$_2$ &\quad\qquad\qquad\qquad $\{1,(0,1],1\}$ &\quad\qquad\qquad\qquad{FM}\\
S$_3$ &\quad\qquad\qquad\qquad $\{1,-1,1\}$ &\quad\qquad\qquad\qquad{AFM}\\
~~~~~~&\quad\qquad\qquad\qquad $\{1,(-1,0),1\}$&\quad\qquad\qquad\qquad{FIM}\\
S$_4$ &\quad\qquad\qquad\qquad $\{-1,0,/\}$ &\quad\qquad\qquad\qquad{AFM}\\
S$_5$ &\quad\qquad\qquad\qquad $\{-1,1,-1\}$ &\quad\qquad\qquad\qquad{AFM}\\
~~~~~~&\quad\qquad\qquad\qquad $\{-1,(0,1),-1\}$&\quad\qquad\qquad\qquad{FIM}\\
S$_6$ &\quad\qquad\qquad\qquad $\{-1,-1,-1\}$ &\quad\qquad\qquad\qquad{AFM}\\
~~~~~~&\quad\qquad\qquad\qquad $\{-1,(-1,0),-1\}$ &\quad\qquad\qquad\qquad{FIM}\\
S$_7$ &\quad\qquad\qquad\qquad $\{1,(0,1],-1\}$ &\quad\qquad\qquad\qquad{FIM}\\
S$_8$ &\quad\qquad\qquad\qquad $\{1,[-1,0),-1\}$ &\quad\qquad\qquad\qquad{FIM}\\
\hline
\bottomrule[1.0pt]
\end{tabular}
      \end{threeparttable}
\end{table}

Corresponding to the momentum-space Hamiltonian~(\ref{H_EB}), the real-space tight-binding Hamiltonian is described as
\begin{equation}\label{H_TB}
\begin{split}
H_{\rm TB}=&-t\sum_{{{\langle
i,j\rangle}_{\parallel}}\alpha}c_{i\alpha}^\dag
c_{j\alpha}-\gamma\sum_{{{\langle
i,j\rangle}_{\bot}}\alpha}c_{i\alpha}^\dag c_{j\alpha}\\
&+\sum_{i\alpha}M_ic_{i\alpha}^\dag s_zc_{i\alpha}
+\sum_{i\alpha}\Delta_ic_{i\alpha}^\dag c_{i\alpha}\\
&+U\sum_{i\alpha}\mu_ic_{i\alpha}^\dag c_{i\alpha}+H'\\
H'=&\frac{i}{3}\sum_{\langle i,j\rangle,\alpha,\alpha'}\lambda_R^{ij} c_{i\alpha}^\dag [({\bm s}\times{\bm d}_{ij})\cdot{\bm \hat{z}})]c_{i\alpha'}\\
&+\frac{i}{3\sqrt3}\sum_{\langle\langle i,j\rangle\rangle,\alpha}\lambda_I^i\nu_{ij}c_{i\alpha}^\dag s_zc_{j\alpha}
\end{split}
\end{equation}
Here, $c_{i\alpha}^\dag$ creates an electron with spin polarization $\alpha$ at site $i$, $\langle i,j\rangle$ ($\ll i,j\gg$) runs over all the (next) nearest-neighbor-hopping sites, and the subscript ${\|}~(\perp)$ means in-plane (out-of-plane), $\alpha$ denotes the real spin index, $\mu_i=+1~(-1)$ holds if site $i$ locates on the bottom (top) layer, ${\bm d}_{ij}$ is the unit vector pointing from site $j$ to site $i$, ${\bm \hat{z}}$ is the $z$-axis unit vector, and $\nu_{ij}=+1 (-1)$ is valid for the clockwise (anticlockwise) hopping. The velocity parameter in Hamiltonian~(\ref{H_EB}) is determined by $\upsilon=\sqrt3at/2\hbar$, where $a=2.46~{\AA}$ is the lattice constant. The parameter $M_i$ takes the value of $M$ when atom $i$ is located on sublattice A$_1$, and the value of $\beta M$, $\eta M$, and $\zeta \beta M$ when atom $i$ is located on sublattice B$_1$, A$_2$, and B$_2$, respectively. The parameter $\Delta_i$ is $\Delta$ if atom $i$ locates on sublattice A$_1$ or B$_1$, and $\eta' \Delta$ if atom $i$ locates on sublattice A$_2$ or B$_2$. By performing Fourier transformation (details are available in Refs.~\cite{McKosh,Neto}), the real-space Hamiltonian~(\ref{H_TB}) becomes the momentum-space Hamiltonian~(\ref{H_EB}).

In particular, note that we disregard other weaker interlayer hopping terms \cite{McKosh} that are not essential for the problem at hand. Typical parameter values $t=2.7$~eV and $\gamma=0.39$~eV \cite{Neto} are used below to perform all the model calculations. At the same time, we choose for the calculations $M=10$~meV, which is much larger than the ignored band trigonal warping effect (about 1~meV \cite{McKosh} near the neutral point for bare BLG). Variation of these parameters affects the band structures quantitatively but does not change the qualitative conclusions.

\begin{figure*}
\centerline{\includegraphics[width=18cm]{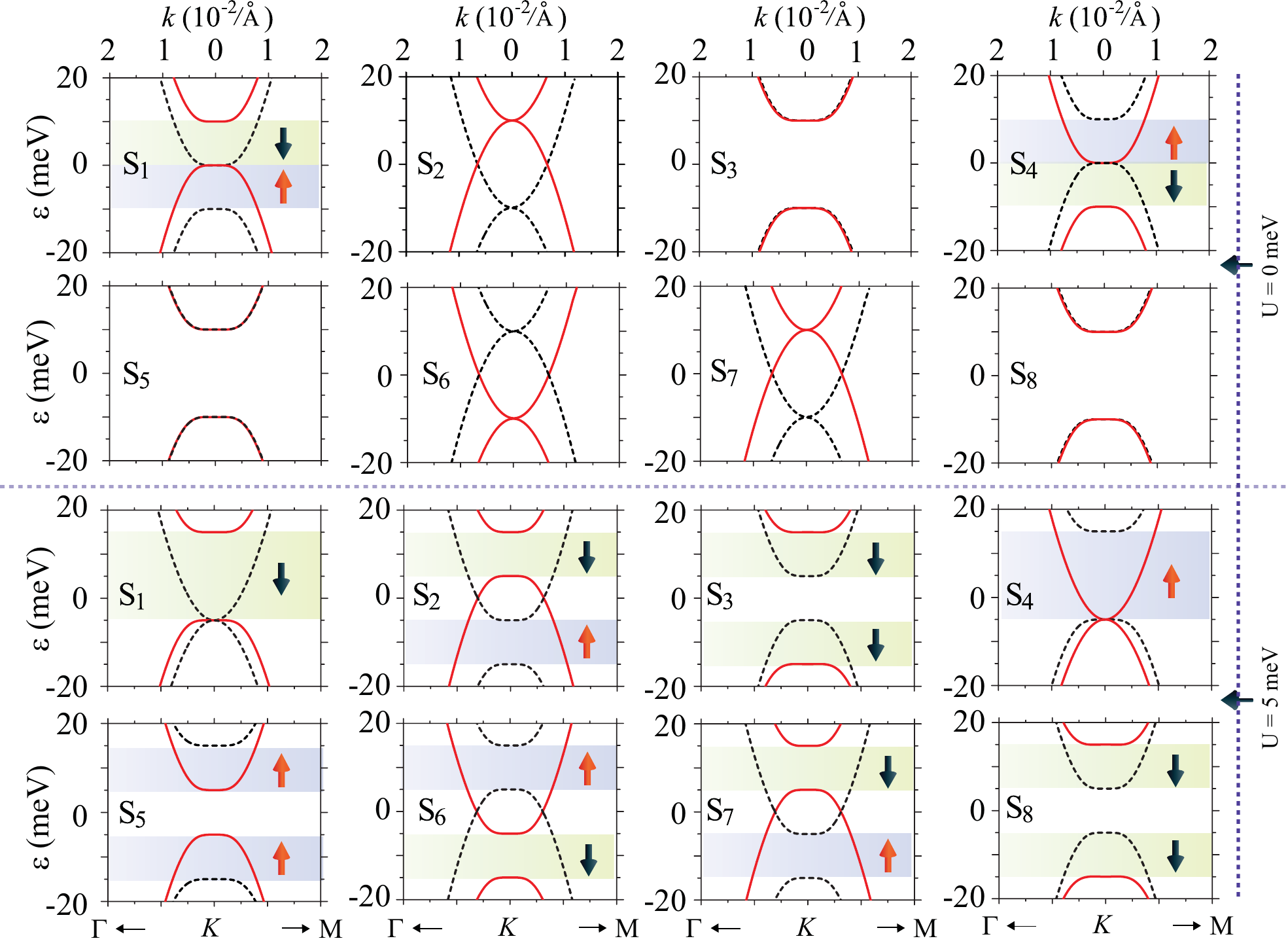}} \caption{Band structures for magnetic states $S_1$ to $S_8$ in Fig.~1(c). The top (bottom) row indicates the results of $U=0$~meV ($U=5$~meV). The red-solid and black-dashed lines correspond to spin-up $(\uparrow)$ and spin-down $(\downarrow)$, respectively. The shaded areas, green for $\uparrow$ and blue for $\downarrow$, represent the regions of complete spin polarization. The parameters $\Delta=0$, $M=10$~meV are fixed. The indices $(\beta,\eta,\zeta)$ take the values of $(1,0,/)$ for S$_1$, $(1,1,1)$ for S$_2$, $(1,-1,1)$ for S$_3$, $(-1,0,/)$ for S$_4$, $(-1,1,-1)$ for S$_5$, $(-1,-1,-1)$ for S$_6$, $(1,1,-1)$ for S$_7$, and $(1,-1,-1)$ for S$_8$.}
\end{figure*}

\begin{figure*}
\centerline{\includegraphics[width=15cm]{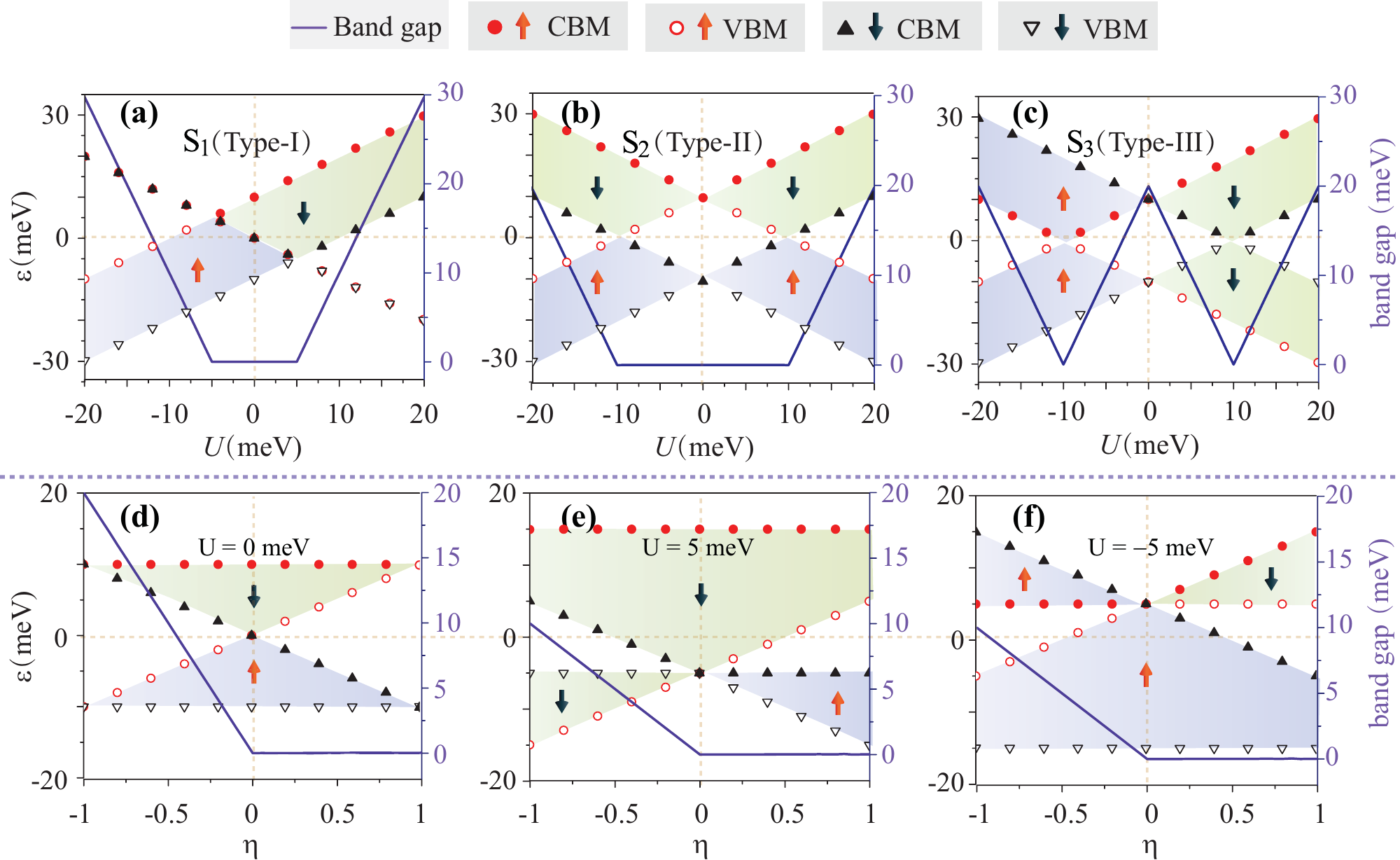}} \caption{(a)-(c) Spin-dependent conduction band minimum (CBM) and valence band maximum (VBM), total band gap as a function of bias parameter $U$ for (a) S$_1$ with $(\beta,\eta)=(1,0)$, (b) S$_2$ with $(\beta,\eta,\zeta)=(1,1,1)$ and (c) S$_3$ with $(\beta,\eta,\zeta)=(1,-1,1)$. (d)-(f) Dependence of spin-dependent CBM and VBM, the total band gap on layer polarization parameter $\eta$ for (d) $U=0$, (e) $U=5$~meV and (f) $U=-5$~meV for $(\beta,\zeta)=(1,1)$. In (a)-(f), the parameters $\Delta=0$ and $M=10$~meV are fixed, and the shadows (blue for spin up and green for spin down) represent the regions of complete spin polarization in the $(\varepsilon,U)$ plane.}
\end{figure*}

\section{Results and discussion}

Since the term $H_\Delta$ in Hamiltonian~(\ref{H_EB}) depends on the environment to which BLG is exposed, we divide the situations into the absence (Fig.~2 and Fig.~3) and presence (Fig.~4 and Fig.~5) of $H_\Delta$ as below. For each situation, we mainly consider the conduction and valence subbands near the Fermi energy, and discuss in depth the spin polarization, spin-dependent band edges as well as the total band gap. Potential device applications are illustrated at the end of this Section. Note that we have disregarded the spin-orbit coupling ${\cal H}'=0$ in this section; we qualitatively discuss its effects below.

\subsection{Absence of staggered sublatttice potential $(H_\Delta=0)$}

In the top panel of Fig.~2, we show the four low-energy subbands for magnetic states $S_1$ to $S_8$ without bias potential ($U=0$~meV). We fix the parameters $\Delta=0$ and $M=10$~meV, and set $|\eta|=1$ and $|\zeta|=1$ if $\eta$ and $\zeta$ are nonzero for simplicity. The results for $U=0$ in Fig.~2 indicate that the spin splitting near the Fermi energy depends strongly on the distribution of magnetism on two opposite layers. For S$_1$ and S$_4$ (or for S$_2$ and S$_6$), although the band structures of the two together can overlap completely, their band spin polarization is the opposite. For S$_2$ and S$_7$, no band difference is observed. For S$_3$, S$_5$ and S$_8$, a band gap is opened, and no spin splitting is observed because all the bands are spin degenerate.

According to the band features shown in Fig.~2, it is reasonable to divide the eight magnetic states ($S_1$ to $S_8$) into three types: S$_1$ and S$_4$ belong to type-I; S$_2$, S$_6$, and S$_7$ belong to type-II; S$_3$, S$_5$, and S$_8$ belong to type-III. This classification depends on the fact that the low-energy four bands near the Fermi energy are basically contributed by sublattices A$_2$ and B$_1$, and the magnetic states in sublattices A$_1$ and B$_2$ are trivial. These three types of situations are distinguishable by the value of $\beta\eta$, which is 0 for the type-I case, 1 for the type-II case and -1 for the type-III case. This can be seen from the model: in the case of interest $\{|U|,|M|,|\Delta|\}\ll\gamma$, the eight-band Hamiltonian~(\ref{H_EB}) can be reduced to the following two-band form in the low-energy regime of $\varepsilon<\gamma$ by treating it perturbatively~\cite{Zhai2022},
\begin{equation}\label{H_FB0}
\widetilde{{\cal H}}_0^s(\bm p)\simeq
\left(\begin{array}{cccc}
-U+s\eta M&-\frac{\upsilon^2}{\gamma}(\pi^\dag)^2\\
-\frac{\upsilon^2}{\gamma}\pi^2&U+s\beta M\\
\end{array}\right),
\end{equation}
which takes $\{\psi_{A_2\uparrow},
\psi_{A_2\downarrow},\psi_{B_1\uparrow},\psi_{B_1\downarrow}\}$ as the atomic basis.
Here, we defined $\pi^\dag=p_x-ip_y$, and $s=+1~(-1)$ denotes spin up (down). Note that the two opposite spin projections are decoupled due to the absence of spin flip.
It is seen from Hamiltonian~(\ref{H_FB0}) that the index $\zeta$ indeed has no influence on the low-energy four subbands, agreeing with the numerical results in Fig.~2, where the difference between S$_3$ and S$_8$ (or between S$_2$ and S$_7$) is indeed invisible.

In the bottom panel of Fig.~2, we show the band results for S$_1$ to S$_8$ by fixing the gate voltage parameter $U=5$~meV. It is shown that the spin splitting for each magnetic state is modulated by gate voltage. For the type-I case, the band region of complete spin-down polarization is increased while that of complete spin-up polarization is reduced. For the type-II case, the band region of full spin polarization is opened. For the type-III case, the total band gap is reduced and the two conduction and valence subbands closest to the Fermi energy have isospin polarization, which is different from the behavior in the type-I and type-II cases. The numerical results are consistent with the analytical expression
\begin{equation}\label{E_0}
\varepsilon_0^s=\frac{\beta+\eta}{2}sM\pm\sqrt{\frac{\upsilon^4 }{\gamma^2}p^4+\left(U+\frac{\beta-\eta}{2}sM\right)^2},
\end{equation}
produced by the diagonalization of Hamiltonian~(\ref{H_FB0}). Here, $\pm$ correspond to the conduction (valence) band.

We further show the influence of gate voltage ($U$) on the band structure in the top panel of Fig.~3, where the information on spin-dependent conduction band minimum (CBM) and valence band maximum (VBM) as well as the total band gap can be extracted. Because the same-type magnetic states (type-I, type-II or type-III) exhibit similar gate-tunable band features, S$_1$ (type-I), S$_2$ (type-II), and S$_3$ (type-III) are chosen as the representative of the same type to perform calculations. By analyzing the evolution of band edges as $U$ varies, we identify the band regions of full spin polarization, as shown by the shaded areas. It is found that the spin splitting near the Fermi energy is always controllable by gate voltage through turning on or off or even flipping spin polarization near the Fermi energy as well as via modulation of the spin-dependent band gap. Two band features are identified: (i) gate voltage induces the electron-hole asymmetry for S$_1$ (type-I), in sharp contrast to S$_2$ (type-II) and S$_3$ (type-III); (ii) the opposite gate voltages ($\pm U$) do not change the energy band structure for S$_2$ (type-II) but reverse the orientation of band spin polarization for S$_3$ (type-III).

In the bottom panel of Fig.~3, we show the influence of the layer polarization index $\eta$ on spin-dependent CBM/VBM and on the total band gap. The cases of $U=0,~\pm5$~meV are shown. As $\eta$ varies, the electron-hole symmetry is preserved for $U=0$ but is broken for $U=\pm5$~meV. It is found that the value of $\eta$ influences the width of the complete spin polarization regions because the difference in magnetic properties between two opposite layers is modified by $\eta$.

Additionally, we argue that both spin-orbit interaction terms in ${\cal H}'$ inevitably affect the band characteristics or even the electronic topological properties  \cite{ZolGmi2018,ZolGmi20,ZhaiXu,ZolFab}. In our latest theoretical work \cite{Zhai2022}, the analytical processes related to spin-orbit coupling and magnetism in BLG are discussed in detail (due to the complexity of the expressions, we do not show them here). In particular, the sign of $\nu$ and $\chi$ in $\cal H'$ is more likely to be negative \cite{Zhai2022,IsLew,ZolFab} due to structural symmetry in real doubly-proximitized BLG samples.

\subsection{Influence of staggered sublatttice potential $(H_\Delta\neq0)$}

\begin{figure}
\centerline{\includegraphics[width=8.5cm]{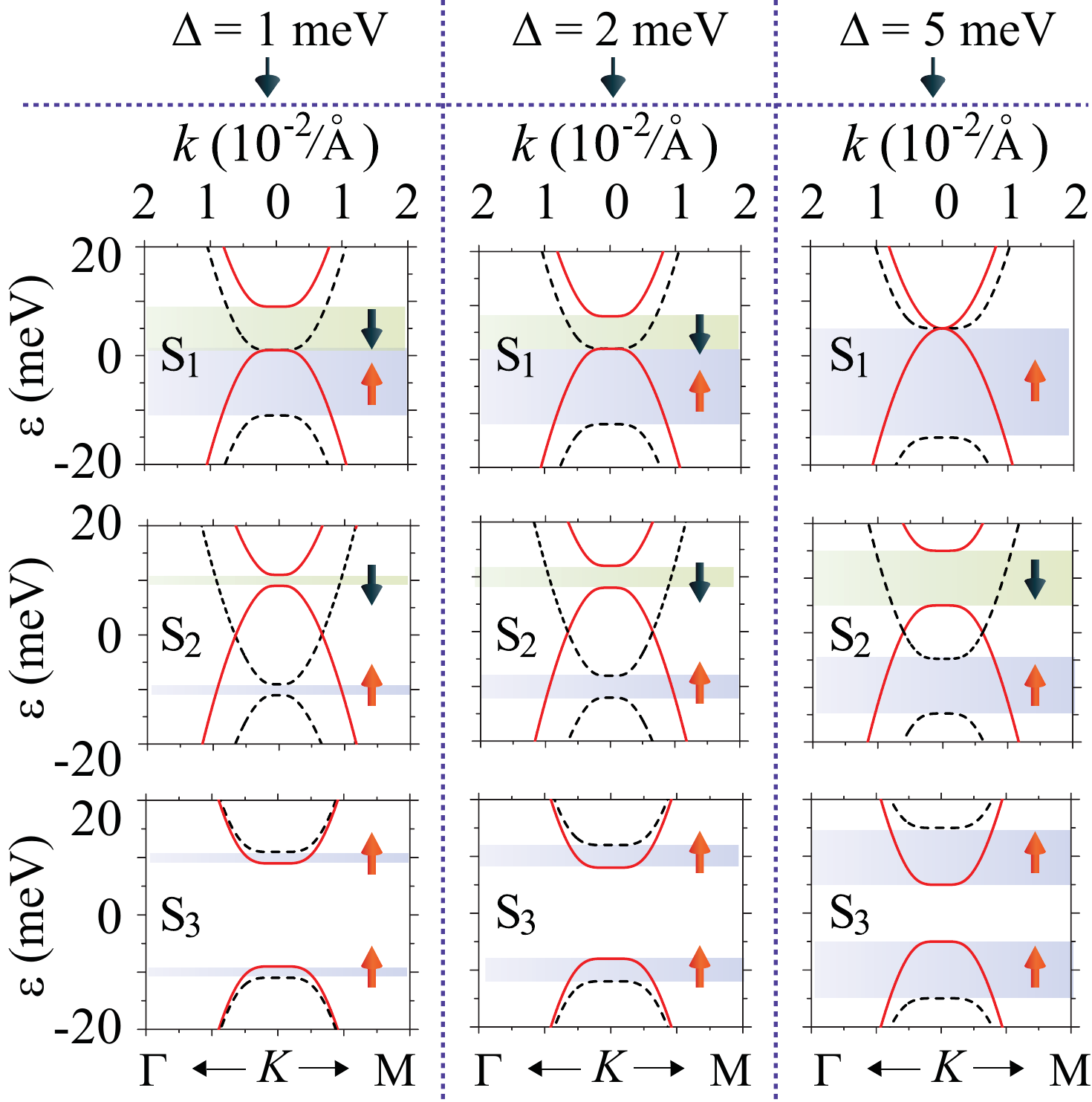}} \caption{(a)-(c) Dependence of CBM and VBM for spin up and spin down and the total band gap on the bias parameter $U$ for S$_1$, S$_2$ and S$_3$ in Fig.~1(c). The parameters take the same values as those used in Fig.~2, except that $\Delta=5$~meV and $\eta'=1$.}
\end{figure}

In Fig.~4, we show the influence of the staggered sublattice potential ($\Delta$) on the band structures (Fig.~2) for S$_1$, $S_{2}$, and $S_3$ without gate voltage. For comparison, the three cases $\Delta=$1~meV, 2~meV and 3~meV are shown. Our results indicate that the band spin polarization is modified by $\Delta$. As $\Delta$ increases from 1~meV to 3~meV, the energy range of full spin-up (spin-down) polarization gets wider (narrower) for S$_1$; the energy ranges of full spin polarizations for both spin-up and spin-down become wider for S$_2$; the total band gap decreases and the energy range of full spin-up polarization becomes larger while complete spin-down polarization is absent for $S_3$. Consequently, the value of $\Delta$ directly affects the energy range in which full spin polarization is observed.

Since the case of spin polarization should be revised according to the environment to which BLG is exposed, the corresponding spin-decoupled two-band Hamiltonian revised by $\Delta$ reads
\begin{equation}\label{H_eff}
\widetilde{{\cal H}}_s(\bm p)\simeq\widetilde{{\cal H}}_0^s(\bm p)+
\left(\begin{array}{cccc}
\eta'\Delta&0\\
0&-\Delta\\
\end{array}\right).
\end{equation}
Here, we can split the second term $\widetilde{{\cal H}}_\Delta$ into two terms: one is
the Fermi-energy correction term $(\eta'\Delta-\Delta)/2$ and the other is $(\eta'\Delta+\Delta){\widetilde\sigma}_z/2$, where ${\widetilde\sigma}_z$ is the $z$ component of Pauli matrix in the \{B$_2$, A$_1$\} sublattice space.
When $\eta'=-1$ holds, $\Delta$ merely shifts the Fermi energy.

By fixing the staggered sublattice potential $\Delta=5$~meV and the index $\eta'=1$, we further show the influence of $U$ on the spin-dependent VBM and CBM as well as the total band gap in Fig.~5. After analyzing the evolution of band edges as $U$ changes, we identify the energy ranges of bands with full spin polarization, as shown by the shaded areas. Comparing the results of Fig.~5 and Fig.~3 (top), we see that the data in Fig.~5 is the result of all the data in Fig.~3 being shifted by 5~meV to the right along the $U$ axis. In this sense, $\Delta=5$~meV is equivalent to the gate voltage $U=-5$~meV. The band structure corresponding to the Hamiltonian~(\ref{H_eff}) is
\begin{equation}\label{Band}
\begin{split}
\varepsilon^s=&\frac{\beta+\eta}{2}sM+\frac{\eta'-1}{2}\Delta\pm\sqrt{\frac{\upsilon^4 }{\gamma^2}p^4+\Omega^2},\\
\Omega=&\frac{\beta-\eta}{2}sM-\frac{\eta'+1}{2}\Delta,
\end{split}
\end{equation}
agreeing well with the above numerical results.

\begin{figure}
\centerline{\includegraphics[width=8.5cm]{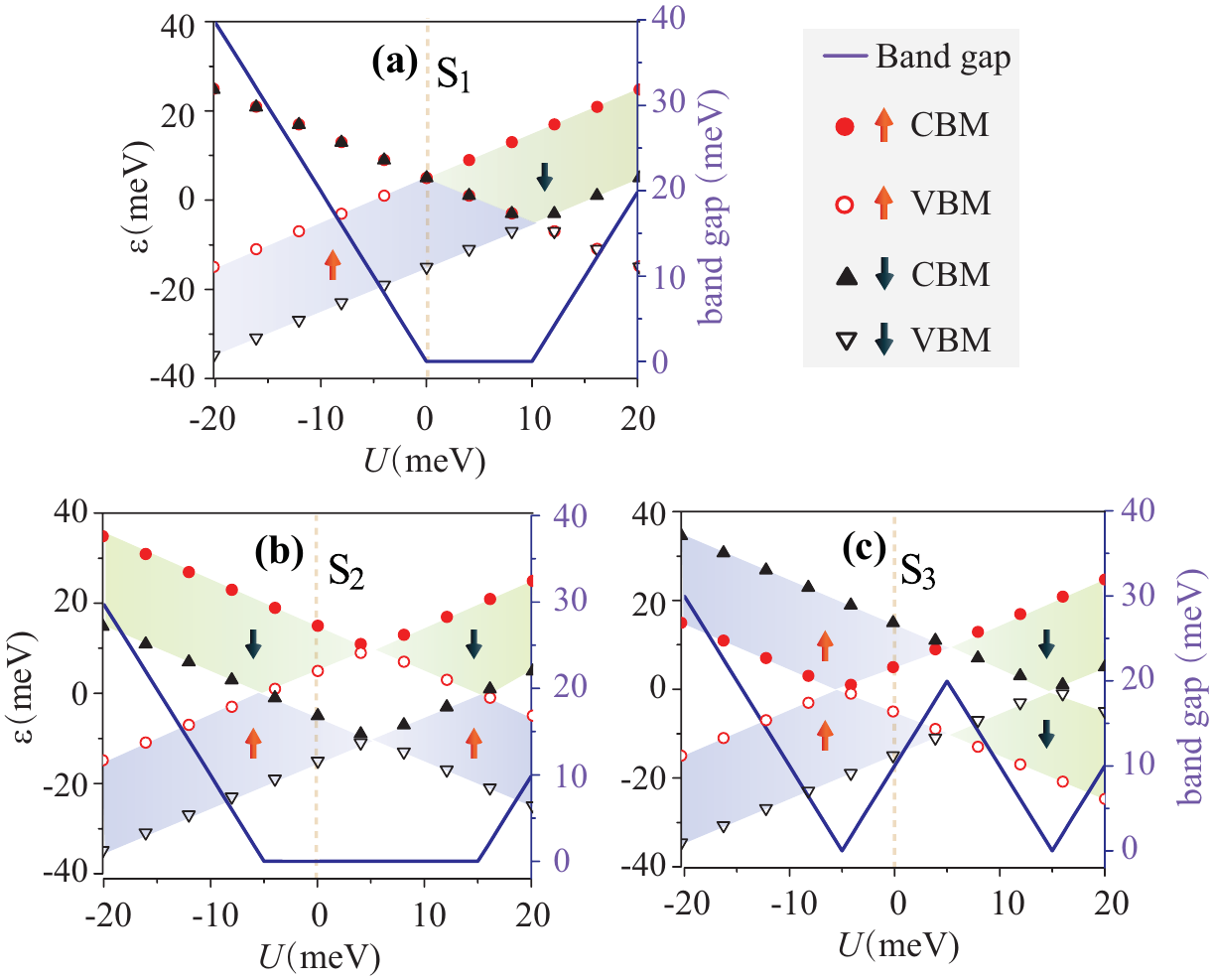}} \caption{(a)-(c) Dependence of CBM and VBM for spin up and spin down and the total band gap on bias parameter $U$ for S$_i$ in Fig.~1(c). The parameters take the same values as those used in Fig.~2, except that $\Delta=5$~meV and $\eta'=1$.}
\end{figure}

We also remark here that the band structure of bilayer graphene is very sensitive to the stacking order, which determines the structural symmetry. Specifically, the case of AA-stacking (inversion and mirror symmetry) has higher symmetry than that of Bernal-stacking (inversion symmetry) in terms of the spatial structure, leading to a parabolic dispersion for Bernal stacking but a Dirac dispersion for AA-stacking \cite{RozSbo}. Nevertheless, we expect that the phenomena related to electrically-controllable spin polarization would be qualitatively similar for Bernal and AA-stacking situations. Quantitative treatment of the band structure for the AA-stacking would require an inclusion of strong Coulomb interaction \cite{RakRoz}.

\subsection{Potential device applications}

Our results in Figs.~2-5 reveal that each magnetic state, $S_1$ to $S_8$, in Fig.~1(c) is capable of exhibiting the complete band spin polarization by tuning gate voltage. This offers a multitude of options for designing spintronic devices by making full use of the layer degree of freedom in BLG. We quantitatively investigate some intresting situations below.

By using quantum transport theory \cite{ZhaiXu,Neto}, the zero-temperature conductance in a two-terminal device is expressed as
\begin{equation}\label{conductance}
G=\frac{e^2}{h}{\cal T}(\varepsilon),
\end{equation}
where ${\cal T}(\varepsilon)={\rm Tr}({\cal
G}\Gamma_{\rm 1}{\cal G}^\dag\Gamma_{\rm 2})$ is the transmission coefficient, $\cal G$ is the
Green's function of the system, Tr denotes the trace, and $\Gamma_{\rm
1, 2}=i(\Sigma_{\rm 1, 2}-\Sigma_{\rm 1, 2}^\dag)$ is the
broadening of two electrodes.

\begin{figure}
\centerline{\includegraphics[width=8.5cm]{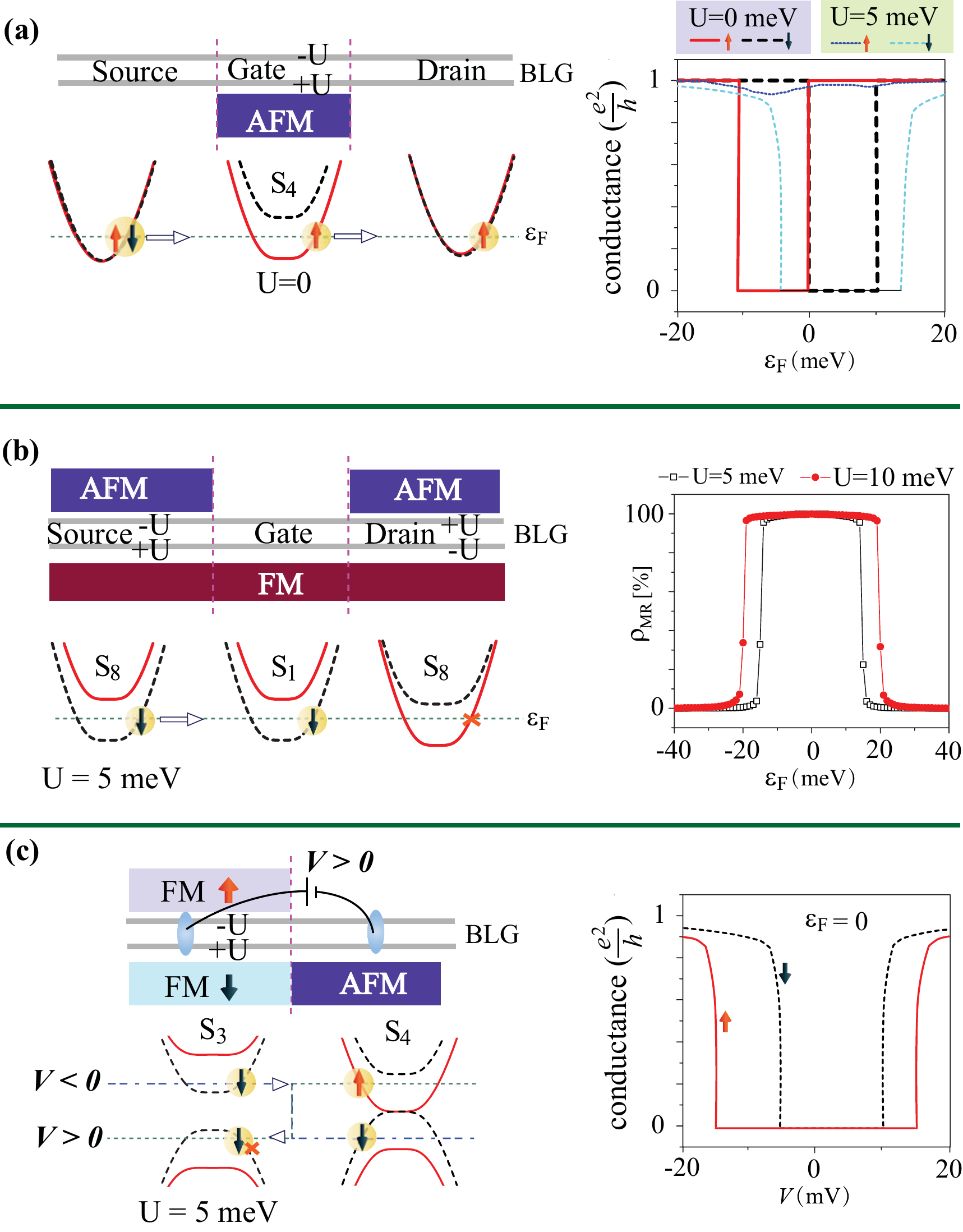}} \caption{Proposed spintronic devices for (a) spin filtering effect, (b) giant magnetoresistance effect, and (c)  spin diode effect under the mechanism of band-to-band spin-matching tunneling. Among the many choices in Fig.~1(c), only S$_4$ in (a), S$_1$+S$_8$ in (b), and S$_3$+S$_4$ in (c) are taken as examples. Here, $\varepsilon_{\rm F}$ denotes the Fermi energy, and ''$\times$" indicates no transmission. The gate voltage parameter $\pm U$ in (a)-(c) can be induced by dual gates \cite{ShiYam}, and $V$ in (c) is a longitudinal bias. The right panels in (a) and (b) show the calculated results for an armchair-edged BLG, for which the width in (a)-(c) is set as 9.84~nm and the gate-region length in (a) and (b) is fixed at 8.52~nm [a sharp interface \cite{Neto} is considered for simplicity in (c)], while other parameters are consistent with Fig.~2.}
\end{figure}

In the left panel of Fig.~6(a), the spin filtering effect is realized in a normal/S$_4$/normal junction, where ``normal" represents the normal bare BLG. Such a junction can be experimentally prepared by locally contacting a bare BLG sample with an AFM material. Due to the short-range property of magnetic proximity effect, it is reasonable to assume that the top layer has no magnetism, while the bottom layer has the spatial nonuniform magnetism. The layer-contrasting mechanism here leads to the result that spin-up electrons take part in conduction while transport of spin-down electrons from the source (electrode 1) to the drain (electrode 2) is blocked for $U=0$ since the band-to-band tunneling must be spin matching. The conductance in this device is shown in the right panel of Fig.~6(a). For $U=0$, spin up is conductive but spin down is blocked for $\varepsilon_{\rm F}\in(0,10~{\rm meV})$, while spin down is conducting but spin up is blocked for $\varepsilon_{\rm F}\in(-10~{\rm meV},0)$, agreeing with the band analysis under the selected parameters. For $U=5$~meV, spin up is conducting but spin down is blocked for $\varepsilon_{\rm F}\in(-5~{\rm meV},15~{\rm meV})$, because $U$ modifies the band spin polarization in the central scattering region (gate).

In the left panel of Fig.~6(b), an S$_8$/S$_1$/S$_8$ junction is investigated as a giant magnetoresistance device. This junction can be prepared by contacting the whole bottom layer of a BLG sample with a FM insulator and at the same time contacting the two electrodes on the top layer with an AFM insulator. Based on Eq.~(\ref{conductance}), the electrically-controllable MR effect is characterized by the ratio \cite{ZhaiXu}
\begin{equation}
\rho_{\rm MR}=\frac{R_{\rm AP}-R_{\rm P}}{R_{\rm AP}+R_{\rm P}},
\end{equation}
where $R=1/G$ is the resistance, and the subscript AP (P) indicates the antiparallel
(parallel) alignment of the gate voltages on both electrodes. The dependence of
$\rho_{\rm MR}$ on the Fermi energy is plotted in the right panel of Fig.~6(b), where we used the values $U=5$~meV and 10~meV. In this case, the AP (P) configuration corresponds to the high (low) resistance state. Determined by the band-to-band spin-matching tunneling mechanism, nearly 100\% magnetoresistance effect is observed for $\varepsilon_{\rm F}\in(-15~{\rm meV},15~{\rm meV})$ at $U=5$~meV and for $\varepsilon_{\rm F}\in(-25~{\rm meV},25~{\rm meV})$ at $U=10$~meV.

In Fig.~6(c), we illustrate that a spin diode effect --- a fully spin-polarized current which is flowing in the forward direction and blocked in the backward direction (or vice verse) \cite{ZhaiWen,ZhaiZhang} --- can be realized in an S$_3$/S$_4$ junction. This junction can be prepared in a BLG sample that is doubly proximitized by two reverse-magnetized FM insualtors (one is at the top, the other is at the bottom) on the left region of BLG and by one bottom AFM insulator on the right region of BLG. For the calculations (right panel), we fix $U=5$~meV in the S$_3$ region. We demonstrate that, for the longitudinal bias 5~mV~$<|V|<10$~mV, spin down is conductive for negative bias but blocked for positive bias, while spin up is always blocked, agreeing well with the spin-matching tunnelling analysis of spin diode effect (left panel). For $|V|<5$~mV, no current is generated for any spin direction due to the presence of band gap in the S$_3$ region. For $|V|>10$~mV, the spin diode effect disappears because more subbands participate in transport. Specifically, the result of $U=0$ in Fig.~6(c) is not plotted here because the spin diode effect is absent.

Note that Fig.~6 shows just three examples of spintronic devices based on layer-contrasting magnetism in BLG. In principle, the spin-filtering effect or the giant magnetoresistance effect can be realized for many combinations from S$_1$ to S$_8$ under the control of gate voltage, revealing the appeal of {\it layered spintronics} --- by tuning the layer degree of freedom we can design devices with various operational principle.

To develop layered spintronics based on the nonmagnetic BLG or other possible 2D materials \cite{LiuHuang,NovMish,ChoiYun,GoZha,LiYang}, the key is to find its suitable magnetic-proximity partners which do not significantly affect the electron structure of the parents. Despite the existence of many early reports in Table~I, a lot of efforts are still needed in the near future, specifically in the exploration of room-temperature magnetism. Beyond the nonmagnetic 2D materials, looking for intrinsic vdW magnets or heteromagnets with high Curie or N\'{e}el temperature \cite{BleiLado,ZhangLu} provides an alternative way to develop layered spintronics.

\section{Conclusions}

In this Article, we have shown how layer-dependent magnetism, induced in bilayer graphene via the magnetic proximity effect when it is brought in contact with various magnetic materials on one or both sides, affects the electron band structure. We considered the cases of ferro-, antiferro-, and ferrimagnetism. In particular, we have discovered that the spin polarization near the Fermi energy depends strongly on the magnetic configuration of BLG. Significantly, the spin polarization can be tuned by gate voltage which shifts the edges of spin-dependent bands and alters the total band gaps. In addition, the spin polarization can be modified by the proximity-induced staggered sublattice potential.

Our main assumption here is that the magnetic structure of the graphene layer in contact with magnetic material (with possible further restrictions as only certain magnetic planes, such as FM Fe(111) plane in AFM BiFeO$_3$ \cite{WuSong,QiaoRen} could be efficient) repeats this magnetic material, or, in other words, that the magnetic proximity effect is (i) complete and (ii) does not cause back-action and does not modify the magnetic structure of the proximity material due to the presence of graphene. Existing experimental studies \cite{GhiKav} are consistent with these two assumptions, however, more research is needed to understand their limitations and the role of back-action. We furthermore disregarded the influence of dipole-dipole interactions on proximity-induced magnetism. We know that in 2D vdW magnets the situation can be even more complicated. For example, FePS$_3$ is a layered antiferromagnet with zigzag ordering (see e.g. Ref. \cite{Zhang21}), and for emergence of the zigzag order second- and third-nearest-neighbor exchange interactions are essential. It would be interesting to expand our research to include other types of interactions, relevant to a broader class of vdW materials.

Based on our findings, in particular, of the gate-voltage controlled regimes of full spin  polarization, we proposed specific devices based on bilayer graphene --- spin filter and giant magnetoresistence, which are fully electrically controlled. They make a full use of the layer degree of freedom, with the advantage of easiness of operation compared with the magnetic field control. This is only the beginning of the research on layered spintronics, and we expect that further research will uncover other interesting and useful effects and applications.

\section*{Acknowledgments}
This work was supported by the National Natural Science Foundation of China with Grants No.~12074193 and No.~61874057.

\end{document}